# A Holistic Framework for Safeguarding of SMEs: A Case Study

Nefeli Bountouni
*Suite5 Data Intelligence Solutions Ltd.*
Limassol, Cyprus
nefeli@suite5.eu

Sotiris Koussouris
*Suite5 Data Intelligence Solutions Ltd.*
Limassol, Cyprus
sotiris@suite5.eu

Alexandros Vasileiou
*Ubitech Ltd.*
Athens, Greece
avasileiou@ubitech.eu

Stylianos A. Kazazis
*Ubitech Ltd.*
Athens, Greece
skazazis@ubitech.eu

*Abstract*— The rapid digitalisation of SMEs, further expedited as a business continuity measure against Covid19 impact, has brought along major cybersecurity challenges, as it creates a fertile landscape for malicious actors, that want to capitalise on the insufficient cybersecurity planning and preparedness of SMEs to conduct low-effort, lucrative attacks. This paper constitutes a case study on the cybersecurity challenges, specificities and the safeguarding of the ATracker, a real-life data collection and analytics engine developed by the SME Suite5. The ATracker has been successfully protected against attacks in conjunction with the PUZZLE Framework, a holistic policy-based cybersecurity solution, addressing major cybersecurity pillars and leveraging on the latest scientific advancements in cybersecurity research.

## I. INTRODUCTION

### A. Motivation

With Small-Medium Enterprises (SMEs) comprising 99% of businesses in Europe and providing employment to more than 100 million people [1] their vital role in EU's growth and social system is apparent. A massive wave of SME digitalisation followed the Covid19 pandemic; from low-scale website improvements, up to the adoption of cloud services, uptake of remote work and the implementation of e-commercial systems, various actions were taken as a countermeasure towards the survival of businesses when massive restrictions were in force [2]. However this rushed transition, came at the cost of creating a wide attack surface for malicious actors, that view SMEs as an easy target due to insufficient planning and preparedness for cybersecurity. Despite the plethora of documents and standards shaping the current cybersecurity legislative and regulatory landscape within Europe and beyond, such as the Cybersecurity Act [3], the recently introduced Cyber Resilience Act [4], the ISO 27000 family with regularly updated standards, for example 27001:2022 [5] or the various guidelines at EU or National Level [2], the lack of an easily interpretable framework and of distinction between provisions for big-scale versus small and medium enterprises renders compliance and safeguarding of an SME still a challenge. Furthermore, budgetary issues and lack of expertise in SMEs hinder the adoption of costly commercial cybersecurity tools and their appropriate configuration. With the majority of SMEs (over 80%) handling information of medium or high criticality [2], the consequences – reputational, legal, operational and the subsequent financial losses – could be tremendous for an attacked SME. Under this prism, the PUZZLE Framework [6] aims to democratize access to innovative cybersecurity services for SMEs, approaching the safeguarding of an SME's digital infrastructure as a multifaceted task targeting various information security controls [7], spanning from network security to personnel awareness. The integration of the PUZZLE Framework in the frame of the Activity Tracker (ATracker), a proprietary hybrid-cloud Internet of Things (IoT) and activity data analytics platform developed by the data intelligence SME Suite5 enhanced the cybersecurity status of the business infrastructure and showcased the applicability of the proposed framework in a real-life environment as a holistic and easily adopted cybersecurity framework for SMEs.

### B. Paper Structure

The remainder of the paper is structured as follows: Section II provides a background on the cybersecurity landscape relevant to the ATracker engine, thus spanning across cloud, IoT and Machine Learning (ML) & Artificial Intelligence (AI) infrastructures. The design principles and layers of the PUZZLE Framework are briefly presented in Section III, while Section IV dives in the actual protection of the ATracker through experimentation with the PUZZLE Framework. Finally, the conclusions of this work and future steps are presented in Section V.

## II. BACKGROUND

This section provides a brief overview on the attack landscape and proposed solutions for the three areas relevant to the ATracker infrastructure, namely; cloud, IoT and ML/AI systems.

Cloud Services are an efficient and reasonable IT alternative for SMEs, that increasingly shift there their operations and data, utilising various formats of service delivery, such as Platform-as-a-Service (PaaS), Infrastructure-as-a-Service (IaaS), Software-as-a-Service (SaaS), Desktop-as-a-Service (DaaS) and Function-as-a-Service (FaaS), for their purposes [8]. In 2021 the 64% of European SMEs had already adopted cloud-based services [2], while the Europe's digital target for 2030 calling for the uptake of cloud computing services, big data and AI by the 75% of European enterprises [9] paves the way for an even more intensive cloud and AI adoption in the near future. Another trend is the cloud integration of Internet of Things (IoT) technologies, wherein accessible cloud servers provide a means of transmission, storage and management of massive volumes of IoT data [8].

Despite the advantages of this synergy, security and privacy concerns arise for the users' data due to the focus shift of malicious actors towards the cloud as a lucrative target [10]. Most common attacks against cloud-based systems can be grouped under four main classes [11]: 1) network attacks, where the adversary is taking advantage of the requirement of the client to access the cloud remotely [12] to eavesdrop on private information (e.g. man-in-the-middle) or compromise the system through overload (e.g. Denial-of-Service (DoS)) 2) malware attacks, wherein malicious software enters the

This work was funded by the PUZZLE "Towards a Sophisticated SIEM Marketplace for Blockchain-based Threat Intelligence and Security-as-a-Service" EU Research and Innovation Project under Grant Agreement No. 883540 system to disrupt, damage, steal, or in general enforce some other malicious actions 3) untrusted cloud administrator attacks, where the attackers succeed in gaining privileged access to the cloud towards achieving their ultimate purposes, such as information theft, tampering analytics, disrupting operations among others and finally 4) data manipulation attacks, where the attackers alter the data or metadata of the users or the system to cause execution failures or other system corruption. At the same time the rapid expansion of IoT solutions is not accompanied by the corresponding adoption of adequate defensive strategies to respond against common IoT threats and weaknesses such as malware, exploitation, poor device management and configuration, insecure transmission protocols, data leakage, plaintext data at-rest and in-transit and more [13]. Machine Learning methodologies are designed and executed under the assumption of a trusted and benevolent environment, thus providing an exploitable advantage to attackers that can manipulate data for their own purposes. Two major attack models against ML are recognised in literature [14]. Poisoning attacks target training data, with the purpose of injecting points that will distort the accuracy of the model [15], while evasion attacks are performed against the ML at testing phase. The ultimate goal of such attacks is the manipulation of the ML system towards specific desirable outcomes, leading to improper decision-making .

Approaches for attack mitigation entail an extensive range of solutions. Software-based solutions constitute a mainstream way of cybersecurity defense, scanning against known vulnerabilities and viruses in order to quarantine the infected assets and isolate the threat within the system. These solutions however have proven insufficient against zero-day attacks, while they are also the first mechanism targeted and inactivated by the attackers once inside the compromised system, in order to obfuscate their activity and continue with the rest of their actions. Learning-based approaches integrated in cybersecurity systems, are seen as a key enabler against common challenges faced by traditional signature-based methods, such as the detection of unknown or advanced attacks, the scale up of security solutions, the protection of dynamic IoT systems [16]. A multitude of learning paradigms - supervised, semi-supervised and unsupervised - have been employed in the widespread study of learning-based cybersecurity [13], but the true use of ML and AI as claimed in commercially available solutions can be disputed and field knowledge is required by the users to ensure that they don't buy an overpriced solution that in core operates as any traditional signature-based system [17]. On the side of hardware-based solutions, such as those based on secure isolated and trusted environments (e.g., enclaves, Trusted Platform Modules (TPMs)) offer a secure key management and execution space that is tamper-proof as it requires physical inference to be broken, but are challenging to implement in cloud-based systems. The alternative of software-based TPMs or composite SGX-TPM implementations is considered, but is still in its early steps [14]. Finally, before reaching the phase of deploying a specific cybersecurity service over an SME's infrastructure , a service discovery, assessment and selection process needs to precede. Commercial platforms hosting individual cybersecurity solutions, such as AWS Marketplace[1] or Cloudflare Apps[2] offer common marketplace features, such as service exploration, categorisation, ratings and contact details. However, they do not support tailored service matchmaking, a feature that would bring added value especially to non-expert users that might face difficulties in selecting and configuring in an optimal way the solutions to guard the complete perimeter of their infrastructures, considering also the highly dynamic nature of the cybersecurity landscape.

### III. THE PUZZLE FRAMEWORK

#### A. Design Approach

PUZZLE approaches cybersecurity through the Security as a Service paradigm [18] offering cybersecurity, privacy and data protection management covering core technical cybersecurity controls [7]. The Framework has been designed in line with the business specificities of SMEs– human, budgetary, technological - and is compatible with the cloud transition of SMEs, without cutting down on the innovatory nature of services. Enhanced user experience, ease of deployment and minimisation of manual intervention allow even non-experts and non-ICT SMEs benefit from the introduced novelties in cybersecurity services, data analytics and sharing. Additionally, it aims to facilitate SMEs in the selection and configuration of the most appropriate services for their infrastructures, based on situational awareness and threat intelligence-based recommendations, thus providing a genuinely dynamic solution tailored to the specific needs of each organisation. With the interplay of results between discrete layers and the visualisation of findings in one place, PUZZLE aims also to overcome challenge of disjoint cybersecurity services that hampers the accuracy of results and the ability of personnel to comprehend the overall status of their infrastructures.

#### B. PUZZLE Layers and Interplay

Conceptually the PUZZLE Framework [6] consists of six discrete layers, comprising the PUZZLE Stack, and acting in synergy, each one offering a different family of services capable to cover the SME under protection, from edge operations up to secure network perimeter and cybersecurity situation awareness (Fig.1).

Namely, the layers comprising the PUZZLE Framework are: Network Analytics layer, an analytics engine deployed in the cloud to perform a holistic security analysis and historical trends analysis on data and alerts produced by the distributed probes;

Fig. 1. The Layers of the PUZZLE Framework [6]

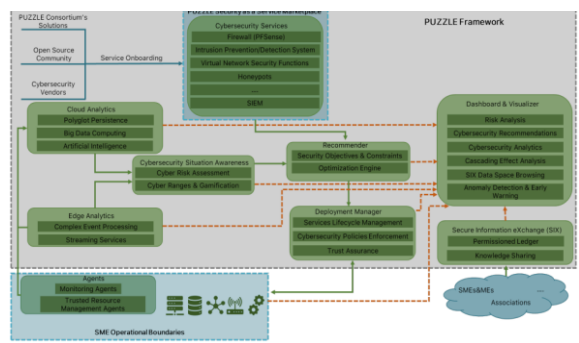

---

[1] https://aws.amazon.com/marketplace/solutions/security  
[2] https://www.cloudflare.com/apps/category/security

Cybersecurity Situation Awareness layer [19], reflecting the security, privacy, trust and operational assurance of the SME through the generation of a risk graph, calculated at design-time and updated at runtime; Edge-Data-Aggregation and Edge-Analytics layer, incorporating agents for network, system and application monitoring, early detection and response to incidents at the edge. Additionally, trust assurance services [20], enhance traffic and cybersecurity service integrity and trustworthiness, while some initial light-weight cybersecurity analytics and calculations are performed at the edge for early response; Secure Information eXchange layer [21], for blockchain-based exchange of threat intelligence in standardized format, for privacy-respecting cross-organisation sharing of new threats to be used by the PUZZLE components but also for human awareness; Recommendation and Security Context Broker, for the suggestion of the most suitable services based on the specific SME's risk assessment, cybersecurity analytics and on the available threat intelligence; Security-as-a-Service Marketplace, constituting the public entry point for the users and organisations to find and utilise services for their private PUZZLE setups, facilitating selection through the interplay with the recommendation mechanism. Finally, a Visualisation Dashboard, receiving data from all layers displays the results of the PUZZLE components to the user and enables interaction.

The PUZZLE Edge and Network Analytics are supported by integrated detection algorithms for behaviour analysis and outlier detection, over encrypted traffic as well as on features extracted from IP/TCP layers (e.g., SYN packets), while ML-based intrusion detection mechanisms are available on the Network Analytic side using Artificial Neural Networks (ANN) for network traffic inspection and feature extraction.

From the aspect of services 'packaging', to allow use by as many SMEs as possible while allowing interoperability with their existing infrastructures, the adoption of a policy-based instead of the traditional executable binary artefacts approach has been favored, to take advantage of the efficient and optimal for cloud-native solutions eBPF packet processing capabilities. The services are available to the SMEs as customisable policy templates addressing a specific attack (e.g. DoS) or offering other functionalities (e.g. vulnerabilities identification, event-based alerting etc.). These templates can be downloaded by the users, edited to match the needs of their SMEs and their organisational and governance rules, and then be onboarded to the PUZZLE stack in order to take effect.

## IV. THE ATRACKER DEMONSTRATOR

### A. ATracker Description and Experimentation Setup

The ATracker is a data collection and analytics solution developed by SUITE5[3]. It comprises three main layers: the ATracker Hubs collecting data from IoT data sources defined by the user. Then these data are consumed by the Personal ATracker (PAT), an application offering data visualisations and simple analytics to Individuals. The third layer is the Cloud-based ATracker (CAT), a cloud-based analytics and visualisation engine that concentrates the data of the various PAT instances at one point where the SUITE5 data analysts can perform combined analytics.

---

[3] https://www.suite5.eu/

Fig. 2. Topology of the ATracker – PUZZLE Demonstrator

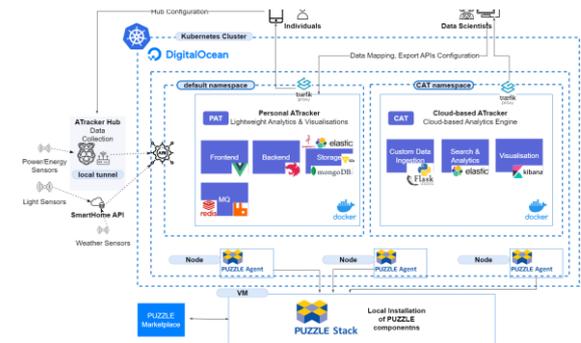

To facilitate experimentation with the PUZZLE framework, a dedicated ATracker deployment has been created, as seen in the topology of Fig.2. The ATracker is hosted in a dedicated Kubernetes cluster with three shared nodes and divided into two namespaces (for the PAT and CAT respectively), while the PUZZLE Stack was installed in a different virtual machine. The connection between the PUZZLE Stack and the Kubernetes cluster was performed in an automated way by adding the necessary information to the infrastructure page of the PUZZLE Dashboard.

### B. Scenarios

The high-level flow of interaction between the ATracker and the PUZZLE Stack is the following. The monitoring data, logs and traffic from the PAT or CAT application and/or namespaces were collected and sent from the PUZZLE agents to the Context Broker, where the aggregated data become available to the rest of the PUZZLE components. The aggregated data were used to identify events based on the applied policies, and trigger afterwards any network-related action as instructed by the deployed policy rule, thus protecting the ATracker application.

An initial proof-of-concept (PoC) of the PUZZLE Framework in action was performed through two use cases: vulnerabilities identification and cybersecurity analytics. Main focus of experimentation was on the establishment of a continuous end-to-end information flow from the ATracker towards the PUZZLE stack, the performance of rule-based cybersecurity analytics and incorporation of user-configured policy templates in the analysis process that would lead finally to the composition of low-level enactment rules enforced to the ATracker.

#### 1) Vulnerabilities Identification and Alerting

This use case is relevant to the identification of vulnerabilities in the overall ATracker infrastructure at design time and the appropriate communication of the results to the ATracker administrator based on severity labelling. The user adds a relevant policy to the Application to create an alert if identified vulnerabilities and severity labels match the rule [more than 4 critical] OR [6 high vulnerabilities] OR [10 scores more than 5.3]. Afterwards the PUZZLE components take the necessary actions in order to gather the results and compare them with the identified vulnerabilities labels and scoring. If the rule is matched, then the Security Context broker prepares a report and sends the results to the PUZZLE Dashboard that is responsible for the visualisation of the vulnerabilities report as well as the display of the alert in the

Application timeline. Vulnerabilities are divided at application/namespace level (Cloud-based ATracker and Personal ATracker/default) to the end user through the intuitive Dashboard.

Results: The design-time vulnerabilities scanning was performed for all the components of the CAT and PAT applications. Existing vulnerabilities were identified and appropriately flagged based on their severity, accompanied also by their score. The average time for scanning completion was less than 1 minute, counting from the deployment of a policy under a specific component up to the generation of the results report. An alert was generated whenever the identified vulnerabilities matched the policy rule.

*2) Cybersecurity Analytics at the Edge and Cloud*

In this Use Case the traffic on the 'Edge' and Cloud side of the ATracker is monitored for rule-based cybersecurity analytics for the early detection, mitigation through low-level enactment rules and timely alerting.

The System Administrator instantiates two Brute Force Attack templates (for the PAT and CAT respectively), defining that it will set an alert and block the requesting IP, if it gets an unauthorised access more than 10 times in one minute. A malicious user, performed brute force attacks against the PAT and CAT through pages that include user authentication. The PUZZLE Node Level Agents (NLAs) are continuously monitoring the PAT traffic in the background. The consecutive unsuccessful login attempts are recognised based on the deployed policy, thus the mitigation action is enforced and a report is generated. The System Administrator can see the alert in the application incidents timeline in the Reports section. There she clicks on the specific alert and views in the details of the incident the blocked IP.

Results: The alert was generated at all times a brute-force attack was attempted, in a timely manner. After the identification of the events, the suspicious IP was blocked, as a first mitigation action until the System Administrator becomes aware of the incident and decides on the next actions.

*C. Experimentation Results*

This section presents the results (Table I.) from the integration of the ATracker with the PUZZLE framework, using a validation framework that quantifies deployment, cybersecurity and vulnerabilities aspects and comparing performance of PUZZLE compared to the use of other cybersecurity solutions.

The deployment of the private PUZZLE Stack installation and the configuration of the PUZZLE node level agents was a straightforward process. Additionally, the identification of the appropriate policy templates in the Marketplace, their configuration and onboarding process in the PUZZLE installation configuration. With regards to vulnerabilities identification, 100% asset coverage for the ATracker infrastructure was achieved, while the execution of the vulnerability scan was efficient and did not interrupt the business operation. An alert was generated at all times identified vulnerabilities matched the configured policy. The cybersecurity analytics and mitigation mechanisms of PUZZLE were also proven effective, as all attacks were detected and blocked in a timely manner, based on the defined policies. The alerting system of PUZZLE communicated incidents almost immediately, while the Dashboard allowed the user acquire an overview of the overall ATracker cybersecurity status in seconds thus facilitated for better decision-making and immediate response actions. For users that want to dive in more details per incident or get a better understanding of the results, the relevant extensive information and reports are also available. In terms of situational awareness, the timely alerting and visualisation of events provide useful tools that allow personnel comprehend the status of the infrastructure. The displayed information and experience could be enhanced through the contextualisation of the events, such as their association with specific assets, and the integration of a notification system.

TABLE I. EARLY EXPERIMENTATION RESULTS

| | **Key Performance Indicators** | | |
|---|---|---|---|
| | *Metric* | *With PUZZLE* | *Without PUZZLE* |
| Deployment | Time to setup cybersecurity solution | 1 hour for PUZZLE stack setup | 1 day for mix of solutions |
| Deployment | Time to deploy node level agents | 5 minutes per node | n/a in traditional commercial solutions |
| Deployment | Time to find, configure and onboard cybersecurity services | 4 minutes to find, configure and upload PUZZLE policy | 30 minutes to find, configure and deploy |
| Vulnerabilities Identification | Asset coverage | 100% | 100% |
| Vulnerabilities Identification | Time for scan execution | < 1 minute per component | 3 minutes per component |
| Vulnerabilities Identification | Time to alert upon vulnerability policy match | Almost immediate (seconds) | After scan completion (15 minutes) |
| Vulnerabilities Identification | Time to monitor system-related events | < 1 minute for timeline overview; 5 minutes for detailed report | 5 minutes |
| Cybersecurity | Percentage of packets not detected which match at least one of the detection rules | 0% | 0% |
| Cybersecurity | Percentage of packets not blocked which match at least one of the block rules | 0% | 0% |
| Cybersecurity | Blocked attempts of access of unauthenticated users by the user authentication | 100% | 100% |
| Cybersecurity | Time to respond to a suspicious activity | Almost immediate (seconds) | Almost immediate (seconds) |
| Cybersecurity | Time to receive information regarding a security incident | Almost immediate (seconds) | Almost immediate (seconds) |
| Cybersecurity | Time to monitor network-related cybersecurity events in ATracker | < 1 minute | 5 minutes |

## V. CONCLUSIONS

The PUZZLE Framework brings an innovative cybersecurity solution for the ATracker data analytics engine facilitating the identification of threats and attacks not able to capture and seamlessly address before; with features spanning from enhanced service exploration and contextualized dynamic service recommendation, up to non-interruptive cybersecurity service deployment, operation in the background and mitigation action enforcement. Through the definition of the customisable cybersecurity policies, a protection barricade was created, adapted to match the ATracker deployment, business and operational specificities and capturing detection, mitigation and awareness aspects. Furthermore, the intuitive Dashboard, raised the personnel's situational awareness regarding the ATracker infrastructure but also the bigger threat intelligence landscape for improved response and prevention of possible incidents. This work also showcases the potential of the PUZZLE framework for wide applicability in diverse SMEs' infrastructures.